\documentstyle[preprint,aps]{revtex}
\begin{document}
\draft

\title{Collective patterns arising out of  spatio-temporal chaos}

\author{Shin-ichi Sasa}

\address{Department of Pure and Applied Sciences,
         College of Arts and Sciences, University of Tokyo,
         Komaba, Meguro-ku, Tokyo 153, Japan}

\date{\today}
\maketitle

\begin{abstract}
We present a simple mathematical model in which a time averaged pattern
emerges out of spatio-temporal chaos as a result of the collective action
of chaotic fluctuations. Our  evolution equation possesses spatial
translational symmetry under a periodic boundary condition.
Thus the spatial inhomogeneity of the statistical state
arises  through a spontaneous symmetry breaking.
The transition from a state of homogeneous spatio-temporal chaos to
one exhibiting spatial order is explained by  introducing
a collective viscosity which relates  the averaged pattern
with  a correlation of the fluctuations.
\end{abstract}

\pacs{05.45.+b,47.52.+j,47.54+r}




Macroscopic structures formed in turbulent flow,
such as roll patterns  seen in clouds
and the red spot on Jupiter, are often observed in nature.
These  phenomena have been  explained using a plausible argument
that  an equation describing mean flow is renormalized by turbulent
fluctuations and displays an instability under certain conditions.
Much progress has been made toward justifying this explanation
\cite{Frisch,Tritton},
but  there has been little study of  the origin of the mean flow.
If a system  possesses spatial inhomogeneity, it is easily understood how
mean flow can arise, but this is not the case in which we are interested.
In this paper, we consider the emergence of mean flow
through  a spontaneous symmetry breaking of the statistical state
in  turbulence.


Several recent experiments seem to show the formation of a
mean flow  in turbulence.
In thermo-convective systems, two different types of turbulence,
soft and hard turbulence, can be observed,
depending on the Rayleigh number, and it has been found that hard
turbulence is characterized by the  existence of mean flow \cite{Chicago}.
Because the experiments in question have been done on systems with small
aspect ratios, effects  of  horizontal boundaries  cannot be neglected, and
the spatial translational symmetry is broken externally.  However, the
systems do possess discrete symmetry associated with a spatial parity.
The mean flow then breaks  the parity symmetry spontaneously, that is,
the direction of the large scale circulation flow is determined by the
initial conditions.  We note that
according to recent numerical simulations of
the Boussinesq equation \cite{Toh},
the horizontal boundaries do not play an essential role
in the formation of mean flow.  We thus expect that
mean flow may occur even in the infinite limit of the aspect ratio.
Actually, the emergence of mean flow in convective turbulence in a system
with large aspect ratio (about 1000) has been found in the
electro-hydrodynamic convection of nematics \cite{Sano},
where persistent roll patterns appear in the turbulent state.
In this case, the mean flow can be thought to spontaneously break the
spatial translational symmetry in turbulence.
Additionally, it has been reported  recently that chaotic wave patterns
produced by the Faraday instability display spatially ordered time averages
\cite{Gollub}.
Similar phenomena have also been discovered in Taylor-Coutte flow
\cite{Swinny}  and rotating convection \cite{Ahlers}.


These experimental results suggest that  mean flow and persistent patterns
can appear spontaneously  in various chaotic systems.
We propose to call these  'collective patterns in spatio-temporal chaos',
because their  emergence results from collective action of turbulent
fluctuations, as explained below. This phenomenon should be considered
together with collective
motion in high-dimensional chaos,  which  has been discovered in
spatially extended systems \cite{Chate,Bohr} and globally coupled systems
\cite{Kaneko,Nah}.  We believe that the collective patterns we consider
presently are closely
related to these types of collective motion.


In this paper, we would like to focus on the universal aspects of the
phenomena in question, not system dependent details.  We first present
a simple mathematical model
which exhibits the emergence of collective patterns out of
spatio-temporal chaos. We then explore
how a transition occurs from a state of homogeneous spatio-temporal chaos to
one displaying spatial order by analyzing this model.




Equations of motion in fluid systems are written as
\begin{equation}
\partial_t q^{(i)}+\vec \nabla \cdot \vec J^{(i)}=0,
\label{eqn:con}
\end{equation}
where $q=(q^{(1)}, \cdots,q^{(M)})$ is an $M$-component field defined
in a $D$-dimensional space,  and
the current $\vec J=(\vec J^{(1)}, \cdots ,\vec J^{(M)})$
is a functional  of the field $q$,
(e.g. $q$ is given as  $q=(\vec v, T)$ in  thermo convective systems.)
Eq.(\ref{eqn:con}) is often referred to as a continuity equation
associated with some conservation law. We note, however,  that
$Q^{(i)}=\int dx^D q^{(i)}$ is not a conserved quantity in
a finite system where the surface integral of the current  does not vanish.
Nevertheless, when we focus on a bulk region in a turbulent system whose
fluctuations develop only to a length scale much smaller than the system
size, we can  regard  $Q^{(i)}$ as a nearly conserved quantity.
This property plays an essential role in the formation of collective
patterns, because  large scale dynamics of conserved quantities determine
the long-time behavior  of fluctuations.
Further, due to the fact that the wavelength of collective roll
patterns is of the same order as the role height in convective  systems,
we  conjecture that the formation of  collective roll patterns is  related
to a breakdown of  conservation laws.
%
%
Based on  these considerations, we propose the following one-dimensional
model
equation in order to study collective patterns:
\begin{equation}
\partial_t q^{(i)}+ \partial_x J^{(i)}=\epsilon^2 F^{(i)},
\end{equation}
where $F=(F^{(1)}, \cdots, F^{(M)})$ is a functional of the field $q$,
and $\epsilon$ is a small parameter which measures the extent of the
breakdown of the  conservation properties.
The system size $L$ is assumed to be much larger than the
correlation length of
fluctuations in  spatio-temporal chaos,
and a periodic boundary condition is assumed so as to maintain spatial
translational symmetry.

The simplest equation exhibiting spatio-temporal chaos
with a conserved quantity is the Kuramoto-Sivashinsky (KS) equation,
\begin{equation}
\partial_t v +\partial_x (v^2 + \partial_x v+\partial_x^3 v)=0,
\end{equation}
which has been derived in several contexts, including chemical waves,
front propagation, etc. \cite{Kbook}.
Statistical properties of turbulent solutions of the KS equation
are homogeneous under  periodic boundary conditions
and have been investigated extensively\cite{Sneppen}.
Although it was found that the homogeneity of the statistical state
is broken under a rigid boundary condition \cite{Zaleski},
we do not consider such an external symmetry breaking.
Keeping in mind the guiding principle of modeling mentioned above,
we studied a modified KS equation supplemented with  a small term
which breaks a conservation law. However, as far as we have checked,
 statistical properties seem to remain homogeneous if $F$ is
restricted to a local functional of $v$.
Proceeding to the next step, we considered a two component system
(i.e. $M=2$) in which
a second field is coupled to the KS equation, and found
that  the dynamical behavior is drastically changed.
The model equation we discuss in this paper takes the form
\begin{eqnarray}
\partial_t v +\partial_x (v^2+\partial_x v +\partial_x^3 v)&=&-b v - d u,
\nonumber \\
\partial_t u -D \partial_x^2 u&=& a u +c v-gu^3,
\label{eqn:model}
\end{eqnarray}
where $a$,$b$, $c$, $d$ and $g$ are assumed to be small positive parameters
proportional to $\epsilon^2$. These values  are chosen
so that the trivial solution $u=v=0$ is stable in the absence of
the term $\partial_x ^2 v$,  which produces spatio-temporal chaos.
The condition for this stability is found to be $b>a$  and $ -ab+cd>0$,
and we therefore assume throughout this paper  that
$a=\epsilon^2$, $b=2\epsilon^2$, $c=\epsilon^2$, $d=3\epsilon^2$, and
$g=\epsilon^2$.




Now, we present results of our numerical simulations of
Eq.(\ref{eqn:model}),
where we have used a simple discretization scheme with a spatial
mesh size  $\delta x=1.0$ and time step size $\delta t=0.1$.
This crude numerical scheme is sufficient for our purposes.
(We note that our numerical scheme preserves a discrete spatial
translational
symmetry.) The  values of $\epsilon$ and $L$  were varied
between  $0.1$ and $0.025$  and between $256$ and $2048$, respectively,
because we are interested in the system behavior under
the asymptotic condition $(\epsilon, L) \rightarrow (0,\infty)$.
The diffusion constant  $D$ is treated as a control parameter.
%
%
In Fig. \ref{fig1}, spatio-temporal patterns are displayed  for
different values of $D$, with  $(\epsilon, L) =
(0.1,256)$. In panel (a), small scale fluctuations are seen as
usual in spatio-temporal chaos,  while in panel (b), we can find that
``a coarse-grained periodic pattern'' is superimposed on the small scale
fluctuations. This suggests that  our model can describe a transition to
spatio-temporal chaos with a collective pattern.
%
%
We confirmed quantitatively the existence of  the average pattern
by measuring the quantity $<v(x)>$, where
we have introduced the abbreviated notation
\begin{equation}
<f>= {1 \over T} \int_{T_0}^{T_0+T} dt f(t).
\end{equation}
Here, $T_0$ is some initial, and $T$ is the averaging time.
In order to check whether or not the field $<v(x)>$ has a spatial variation
in the limit $T \rightarrow \infty$, we measured the quantity
\begin{equation}
\sigma^2=
{1 \over L}\int dx \left|<v(x)>-{1 \over L}\int dx\ <v(x)>  \right|^2,
\end{equation}
where the spatio-temporal chaos is considered to be homogeneous
when the scaling relation $\sigma \sim  1/\sqrt{T} $ is obtained
for  sufficiently large $T$.
Figure \ref{fig2} shows the $T$ dependence of  $\sigma$ for
systems in which $(\epsilon, L)=(0.1,1024)$.
$\sigma$ decays to 0 as $\sigma \sim 1/\sqrt{T} $ when $D=1.5$,
while  it seems to keep a constant value when $D=0.5$.
We thus expect that there is a critical value $D_c$ between 0.5 and 1.5.
The $D$ dependence of $\sigma$ for a sufficiently large $T$ $(T=12800)$,
the phase diagram  for our system,
is  shown in Fig. \ref{fig3}.  This  graph suggests that
$D_c$ is near 0.8.


We now attempt to explain  how the collective patterns
emerge out of spatio-temporal chaos.
First, let us derive  a self-consistent condition  under which
a long time averaged quantity $<v(x)>$  forms a periodic pattern.
For simplicity, we focus on the asymptotic case that
$(\epsilon,L) \rightarrow (0,\infty)$ and suppose
the $\epsilon$ dependences of $<u>$ and $<v>$
can be expressed as $<u>=U(\epsilon x)$, and  $<v>=V(\epsilon x)$.
These scaling forms are consistent with numerical results.
Then,  substituting $u=<u>+u'$ and $v=<v>+v'$ into Eq.({\ref{eqn:model}),
averaging the equation, and discarding $o(\epsilon^2)$ terms
in the  equation, we obtain
\begin{eqnarray}
\partial_x(V^2+<v'^2>+\partial_x V)=-b V-dV, \cr
  -D \partial_x^2 U = a U+cV-gU^3,
\label{eqn:avpat}
\end{eqnarray}
where we have assumed  $<(u-<u>)^k>=o(\epsilon)$ with $k=2,3$.
Now, in order to obtain a closed form of the equation for $U$ and $V$,
we  employ the phenomenological relation
\begin{equation}
<v'^2>=\alpha - \nu \partial_x <v>,
\label{eqn:turvdef}
\end{equation}
which has been used to analyze turbulent shear flow \cite{Tritton}.
Here, $\alpha$ and $\nu$ are assumed to be constants, and
$\nu$ is often referred  to as  "turbulent viscosity".
We can see from Eqs.(\ref{eqn:avpat}) and (\ref{eqn:turvdef}) that
$\nu$ expresses how turbulent fluctuations contribute a dissipation
effect on the coarse-grained pattern.
Although the validity of Eq.(\ref{eqn:turvdef}) for our equation
is not proved mathematically, as shown in Fig. \ref{fig4},
our numerical results for $<v'^2>$ and $<v>$
are consistent with this relation.
Further, under  a  mode truncation approximation, we write
$<v>=R\sin(kx+\phi)$ and  $<u>=S\sin(kx+\phi)$,
and obtain the equation for $R$ and $S$
\begin{eqnarray}
(D_{*} k^2 +b)R+d S  &=&0, \cr
 -cR+ (D k^2-a)S &=& -gS^3,
\label{eqn:sce}
\end{eqnarray}
where $k$ is of order $\epsilon$, in accordance with the above stated
assumption, and  $D_{*}=\nu-1$.
Equation (\ref{eqn:sce}) always has the trivial solution $R=S=0$,
which represents a spatially homogeneous statistical state, while
a non-trivial solution  exists only when the inequality
\begin{equation}
D < {2 -\sqrt{3} \over 2} D_{*}
\label{eqn:scc}
\end{equation}
is satisfied.
This result is consistent with our phase diagram shown in Fig.3.
In fact, by fitting our numerical data to Eq.(\ref{eqn:turvdef}),
we found  $\nu \sim 5.5  $  in  the case that
$(D, \epsilon, L)=(0.2, 0.1, 1024)$.
Using this value of $\nu$ for all $D$,
we obtain $D_c \sim 0.6$.
The discrepancy between this value and that obtained from
our phase diagram in
Fig. \ref{fig3} may be due to the fact that $\nu$ depends on
$(D, \epsilon, L)$.


The self-consistent approach developed above can
tell us nothing about the stability of the obtained solutions.
In order to consider the stability problem,
we need to develop a theory describing the slow evolution of
coarse-grained patterns.
A  derivation of  the  evolution equation
from our model equation is beyond the scope of this paper,
but we present the results we obtained using the scaling relation such that
$v(x,t)=V(\epsilon x, \epsilon^2 t)+v^\prime(x,t)$.
We found that the homogeneous statistical state becomes unstable
when the inequality (\ref{eqn:scc}) is satisfied, and in this case,
a coarse-grained periodic pattern with a wavenumber $q_m$  grows
with the maximum growth rate, where
\begin{equation}
q_m=\epsilon\sqrt{ {D_*-2D \over 2D_*D}}.
\end{equation}
That is, a non-trivial persistent pattern appears when the homogeneous
statistical state becomes unstable. This statement is
consistent with our numerical experiments.
Here, we would like to stress that  $D_c$ and $q_m$ are determined by
$D_*$, which is not a  parameter of our model but
associated with  correlations of fluctuations.
Therefore, the existence of the averaged pattern cannot be explained
by developing a linear stability analysis of the trivial state $u=v=0$.


As argued above, the collective action of  fluctuations causes
an instability leading to the formation of persistent patterns.
In this sense, we can say that
fluctuations are the origin  of a spatial order.
On the other hand, one may suspect that
fluctuations act as a turbulent noise, inducing
a random drift of our coarse-grained pattern, like
Brownian motion, and as a result the averaged pattern vanishes
in the limit that the averaging time $T$ goes to infinity.
In order to check this, we calculated
a time series of the complex amplitude of the Fourier mode
associated with the collective pattern,
$\{A(t)\exp(i \phi(t))\}$.
 Spectra of  the amplitude $A$ and phase $\phi$,
which are denoted by $S_A(\omega)$ and $S_\phi(\omega)$ respectively,
are shown in Fig. \ref{fig5}.
The spectrum $S_\phi (\omega)$ goes as
$\omega^{-2}$ at low frequency ($ \omega < \omega_c$).
 This implies the existence of a phase diffusion of
the coarse-grained pattern and therefore that the averaged pattern will
vanish when the averaging time $T$ is chosen much larger
than $2 \pi /\omega_c$.
We should note, however, that  the crossover
frequency $\omega_c$ is much smaller than the inverse of the correlation
time of chaotic fluctuations,
and  from our preliminary numerical results
we believe that  $\omega_c \rightarrow 0 $ for $\epsilon \rightarrow 0$,
with $\epsilon L$ fixed. This  implys that
if the limit $T \rightarrow \infty$ is considered
{\it after} taking $\epsilon \rightarrow 0$  with fixed
$\epsilon L \ge O(1)$,
persistent patterns actually exist.  Figure 5  shows further that
the amplitude $R$ has a long-range correlation.
At low frequency ($\omega < \omega_c$),
the form of $S_R(\omega)$ seems  to be a Lorenzian type, proportional to
$ (1+(\omega\tau)^2)^{-1}$.
Here,  the correlation time  of the amplitude fluctuation, $\tau$,
which may depend on $(\epsilon, L)$, is much larger than $2\pi/\omega_c$.
Although we believe that
$\tau\omega_c  \rightarrow \infty $ for $\epsilon \rightarrow 0$
with fixed $\epsilon L$,
it is necessary to  study the $(\epsilon, L)$ dependence of the system
behavior further.


In conclusion, we have shown that our model equation
can exhibit the emergence of collective patterns out
of spatio-temporal chaos. We believe that
such a phenomenon will be observed in other mathematical models
which have nearly conserved quantities.
By introducing a small parameter $\epsilon$
which measures the extent of the breakdown of  conservation properties,
we have developed  a self-consistent analysis which describes
a transition to the spatially ordered phase.
Both fluctuations of  the amplitude and phase of the Fourier mode
associated with the collective pattern have  long range correlations
at a time scale larger than $2\pi/\omega_c$,  while
over a time scale $2\pi/\omega_c$, which  goes to infinity
for $\epsilon \rightarrow 0$,  averaged patterns are properly defined.
%
%
In our theoretical analysis, the relation between the averaged pattern and
the correlation of fluctuations around it plays a crucial role.
Although  the relation  (\ref{eqn:turvdef}) is a familiar one
at a phenomenological level, there may be  various forms
which relate  the statistical property of fluctuations  with
an averaged quantity. We expect that as yet unknown relations will be
discovered
in experimental systems and other mathematical models.
Also, to determine the correspondence between our model and experimental
systems
may be an important subject. Such studies should contribute greatly
to the understanding of the nature of spatio-temporal chaos.


I am  grateful to  M. Sano for active exchange of ideas concerning his
experiment and the present study.
I thank Y-h. Taguchi and Y. Iba for their critical comments.
I also thank  Y. Kuramoto, S. Toh, S. Nasuno, G.C. Paquette, K. Kaneko
and S. Kai for fruitful discussions.
H. Chat{\'e}, P. Marcq and P. Manneville  are acknowledged
for their hospitality and stimulating discussions during my three week stay
at Saclay in the early stage of this study.



\vskip1cm


\begin{figure}
\caption{
Spatio-temporal patterns of the $v$ field of  our model equation
with $(\epsilon,L) = (0.1,256)$. We have, (a) $D=0.5$ and (b) $ D=1.5$.
The time goes from $t=10000$  to $t=22800$.
}
\label{fig1}
\end{figure}


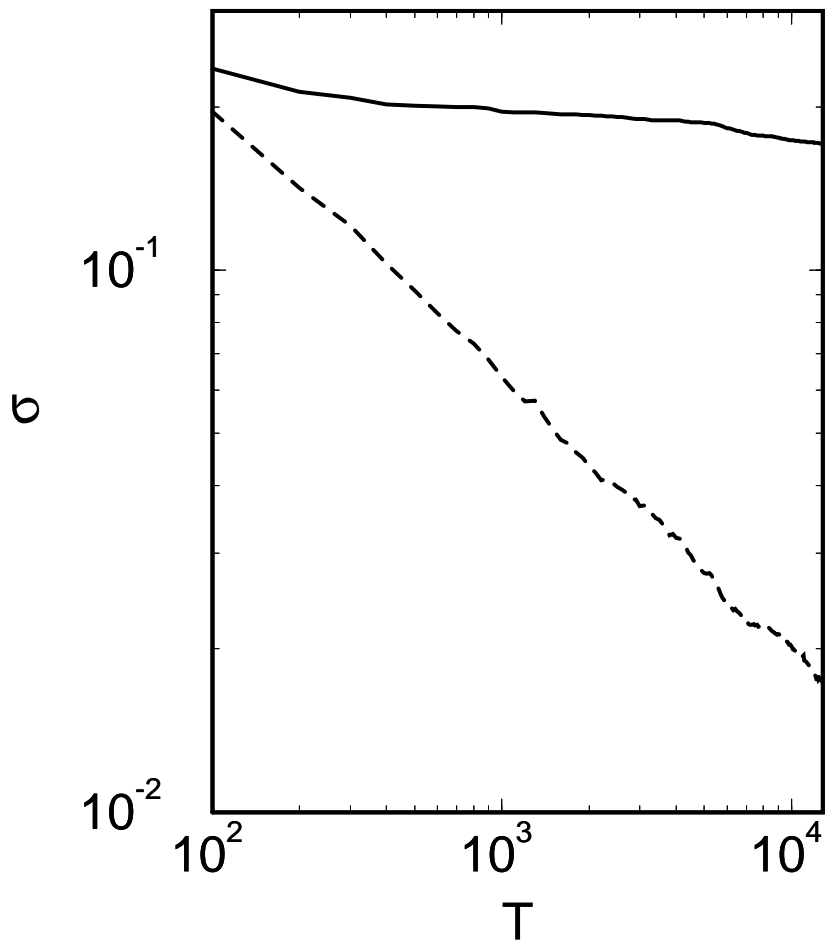
\begin{figure}
\caption{
$\sigma$ versus $T$ for $D=0.5$ (solid line) and $D=1.5$ (dashed line)
$(\epsilon,L) = (0.1,1024)$.
The transient time was chosen as $T_0=10000$.
}
\label{fig2}
\end{figure}


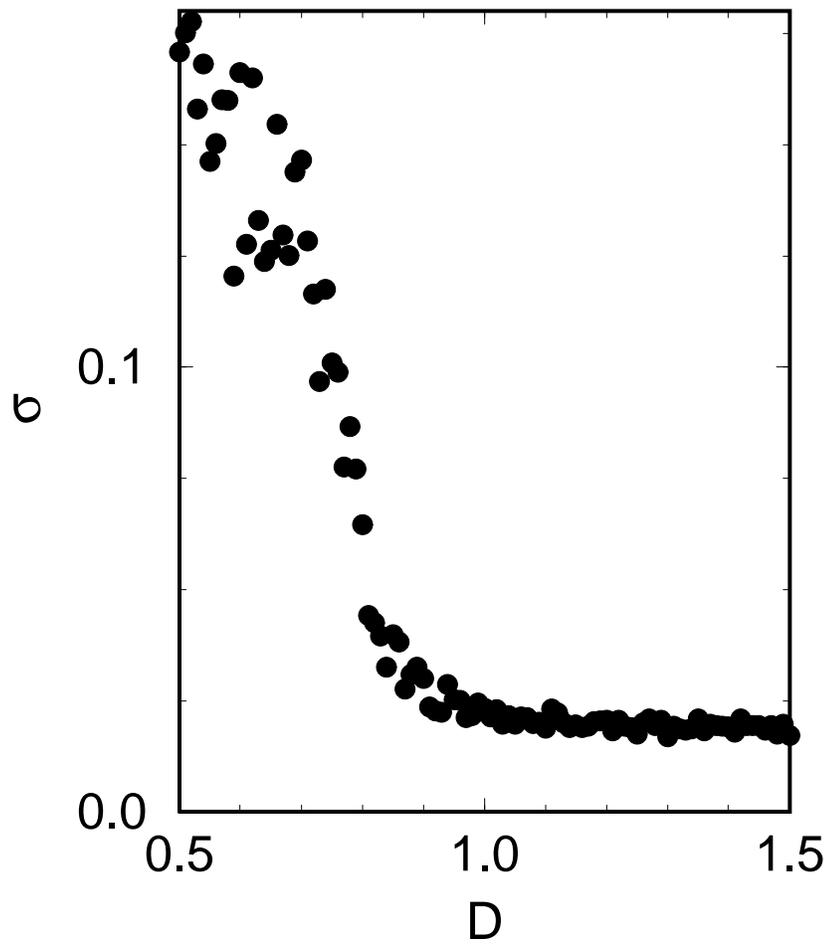
\begin{figure}
\caption{
$\sigma$ versus $D$ for $T_0=10000$ and $T=12800$.
$(\epsilon, L)$ is as in Fig. 2.
Initial conditions are the same for all values of $D$.
}
\label{fig3}
\end{figure}


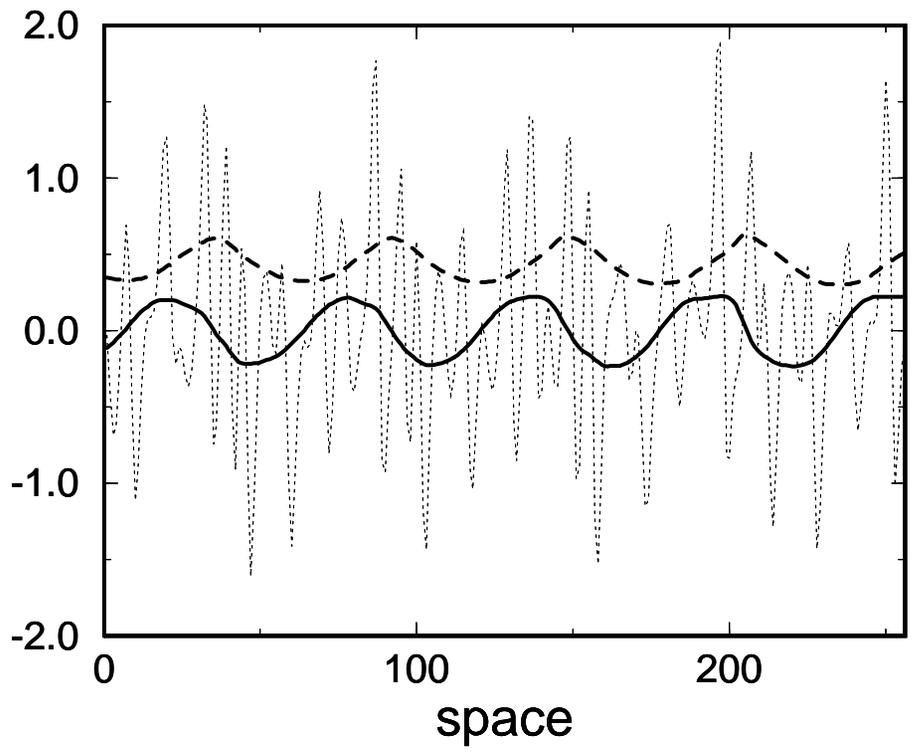
\begin{figure}
\caption{
Snapshot of the $v$ field at $t=505000$ (dotted line),
the averaged pattern (solid line), and
the dispersion at each point (dotted line),
for the system with $(D,\epsilon, L)=(0.2,0.1,1024)$.
Each field in the region $0<x<256$ is displayed.
The statistical procedure was carried out for a sample
$\{v(x,t); 5000< t< 505000\}$.
}
\label{fig4}
\end{figure}


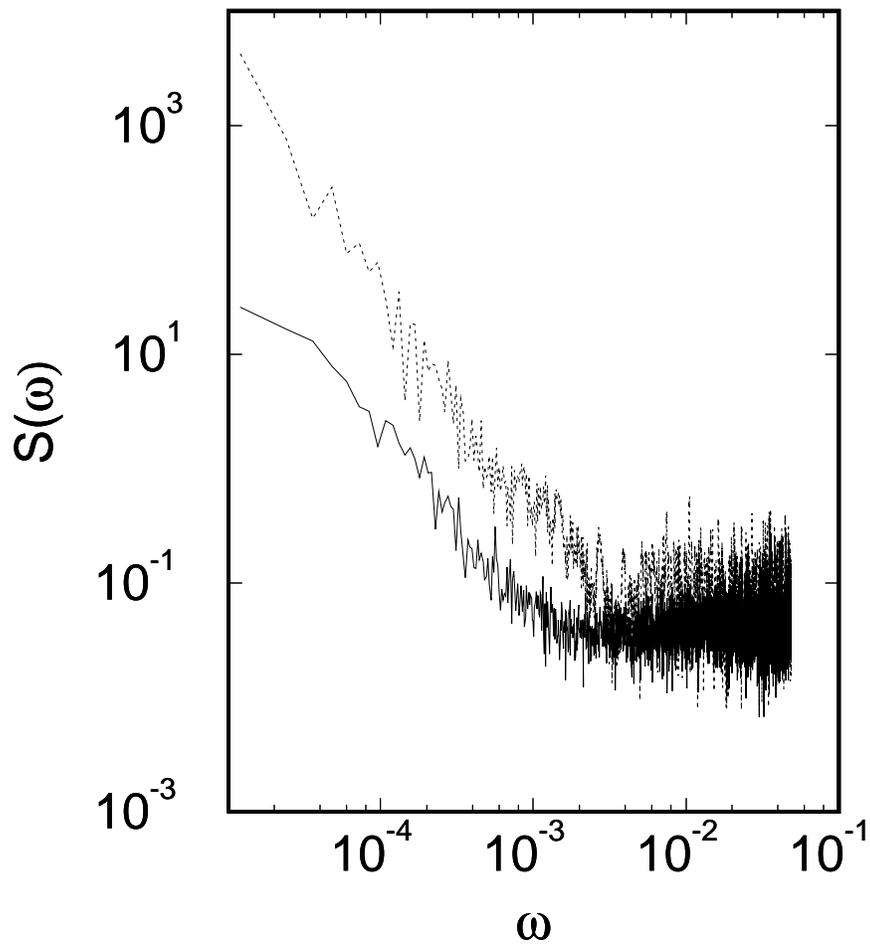
\begin{figure}
\caption{
Spectra for a time series of the amplitude (solid line)
and phase (dotted line) of a Fourier mode $\exp(ikx)$
associated with a collective pattern with the same parameter values
as in Fig. 4. The value of $k$ is given as the position of the peak
of the spectrum of the averaged pattern.
These graphs  were obtained  by averaging 8 samples of the spectra
using 8192 data points  every 128 time units.
}
\label{fig5}
\end{figure}

\end{document}